\documentclass[11pt,a4paper]{article} 
 
\usepackage[english,german]{babel} 
\usepackage{ae,aecompl} 
\usepackage{url} 
 
\usepackage{amsmath,amssymb} 
\usepackage{amsfonts,bbm,theorem} 
 
\usepackage{epsfig} 
\newcommand{\myfigure}[2]{ \includegraphics*[#1]{#2} } 
 
\numberwithin{equation}{section} 
 
\theorembodyfont{\rm} 
\newtheorem{theorem}{Theorem}[section] 
 
\newtheorem{corollary}[theorem]{Corollary}

\newtheorem{proposition}[theorem]{Proposition} 
\newtheorem{remark}[theorem]{Remark} 

\newenvironment{proof}[1][]{\begin{trivlist}\item[]{\em Proof#1: }\/}{%
\hfill\mbox{$\Box$}\end{trivlist}} 
 
\newcommand{\zcrit}{z_{\text{c}}} 
\newcommand{\Nsw}{n_{\text{SW}}} 
\newcommand{\Nhash}{N_{\text{h}}} 
\newcommand{\nm}{n_{\text{M}}} 
 
\newcommand{\dperc}{d_{\text{perc}}}

\newcommand{\defset}[2]{ \left\{ #1: #2 \right\} } 
 
\newcommand{\mean}[1]{\langle #1\rangle} 
 
\newcommand{\rmd}{{\rm d}}

\newcommand{\Z}{{\mathbb Z}} 
\newcommand{\R}{{\mathbb R}}

\newcommand{\norm}[1]{\Vert #1\Vert} 
\newcommand{\order}[1]{{\cal O}\!\left( #1\right)} 
\newcommand{\Aord}{A^{\text{ord}}} 
\newcommand{\Adis}{A^{\text{dis}}}

\def\be{\begin{equation} \label} 
\def\ee{\end{equation}} 
\def\ba{\begin{array}} 
\def\ea{\end{array}}

\def\car{\curvearrowright} 
 
\def\a{\alpha} 
 
\def\g{\gamma} 
\def\G{\Gamma} 
\def\d{\delta} 
\def\D{\Delta} 
\def\e{\varepsilon} 
\def\L{\Lambda} 
 
\def\s{\sigma} 
 
\def\ph{\varphi}

\def\cG{{\cal G}} 
\def\cE{{\cal E}} 
\def\cX{{\cal X}} 
 
\def\bX{{\mathbf{X}}} 
 
\begin{document} 
\selectlanguage{english} 
 
\title{\textbf{
The Continuum Potts Model at the Disorder-Order Transition --\\
a Study by Cluster Dynamics}
} 
 
\author{ 
\small Hans-Otto Georgii\\[0.1cm] 
{\it \small Mathematisches Institut, Ludwig-Maximilians-Universit\"at} \\ 
{\it \small Theresienstr. 39, 80333 M\"unchen, Germany} \\ 
{\small georgii@math.lmu.de}\\ [0.5cm] 
\small J\'ozsef L\H{o}rinczi and Jani M. Lukkarinen\\[0.1cm] 
{\it \small Zentrum Mathematik, Technische Universit\"at M\"unchen} \\ 
{\it \small Boltzmannstr. 3, 85747 Garching bei M\"unchen, Germany} \\ 
{\small lorinczi@ma.tum.de, jlukkari@ma.tum.de} \\ [1cm]}
\date{} 
\maketitle 

\begin{abstract} 
\noindent 
We investigate the continuum $q$-Potts model at its transition
point from the disordered to the ordered regime, with particular emphasis
on the coexistence of disordered and ordered phases in the high-$q$ case.
We argue that occurrence of phase transition can be seen as percolation in the
related random cluster representation, similarly  to the lattice Potts model,
and investigate the typical structure of clusters for high $q$.
We also report on numerical simulations in two dimensions 
using a continuum version of the
Swendsen-Wang algorithm, compare the results with earlier simulations
which used the invaded cluster algorithm, and discuss implications on the 
geometry of clusters in the disordered and ordered phases.

\medskip 
\noindent 
KEYWORDS: Continuum Potts model, first-order phase transition, continuum
percolation, continuum Swendsen-Wang  cluster algorithm 
\end{abstract} 
\vspace{1cm}

\section{Introduction} 
The Potts model is one of the classical models of statistical mechanics
exhibiting a phase transition. In its standard version, it is defined on the
square lattice $\Z^2$, where it was first studied \cite{Potts}. The parameter
controlling its phase transition is temperature: at sufficiently high
temperatures the Gibbs measure is unique, while for low temperatures the
number of translation invariant extremal Gibbs measures (pure phases)
coincides with the cardinality of the spin state space (usually denoted by
$q$) \cite{GHM}. Moreover, when $q$ is large enough, these regimes meet at a
specific value of the temperature where $q$ distinct ``ordered'' phases and
one ``disordered'' phase coexist. This was shown in \cite{KS82} by using
reflection positivity arguments, and in \cite{BKL85,LMMRS91} by
Pirogov-Sinai theory. 
The transition is in this case of first order meaning that exactly at the 
transition point the derivative of the pressure with respect to temperature 
has a jump.
 
The mechanism of phase transition is well understood. When $q$ is in the low
range, the appearance of phase transition is explained by ground   state
degeneracy, similarly to the Ising model: the Ising model has two  ground
states (indeed, it is equivalent with the Potts model for $q=2$),   while the
Potts model has exactly $q$ such states. As $q$ is taken larger, there is ever
increasing possibility for entropy to take domination, and for large enough
$q$ there is a temperature at which energy and entropy balance each other,
leading to coexistence of the high and low-temperature phases.
 
To understand the structure of these pure phases, one has to have a notion
of what a typical spin configuration looks like in each such Gibbs state.
Presently, there is a suggestive picture offered by the random cluster
representation. According to this, phase transition in the Potts model
occurs exactly at that value of the temperature for which the associated
random cluster model is at the percolation threshold \cite{ACCN87,H98}.  This
is possible to prove for $d \geq 2$, and in fact for quite general
underlying lattices. The structure of the percolation clusters allows a good
insight into the typical configurations of the Potts pure phases.
 
In this paper we consider the similar problem of non-uniqueness for the
Potts model placed in the continuum instead of a lattice. 
In this case, the
particles are not sitting in lattice points but at random points in
$\R^d$. The \emph{a priori} distribution of the set of occupied positions is
Poisson process with intensity $z$, which is then modified by the Potts
interaction --- a repulsion between particles of different types.  (In the 
case of zero temperature, this model coincides with the multitype  hard-core
exclusion model of Widom and Rowlinson \cite{WR70}.) The crucial parameter
is then the activity $z$: It is known that, for each number $q$ of different
types and at any fixed temperature, there is only one Gibbs measure whenever
$z$ is small enough, while for high enough values there are $q$ translation
invariant extreme Gibbs measures; see \cite{BKL85,LL72,WR70} and the more
recent 
papers \cite{CCK95,GH96} using random cluster methods.  However, in the
continuum setting it has not been proved so far that these two regimes meet
at a single critical activity $z_c$, and that the phase transition at $z_c$
is of first order when $q$ is large enough.

The difficulty lies in the fact that there are no obvious extensions
to the continuum of the methods available for the lattice Potts
model. On the one hand, reflection positivity does not apply to
continuum models. On the other, the contour techniques of
Pirogov-Sinai theory used in  \cite{BKL85,LMMRS91} do not appear
to admit an immediate extension.
These papers either deal with the continuum Widom-Rowlinson model,
but then only cover the case when $z$ is large and $q$ is held fixed (not
necessarily large),  proving coexistence 
of just $q$ phases, or show the coexistence of $q{+}1$ 
phases at $z_c$ for large $q$, but this only for the \emph{lattice}\/ Potts
model. 

As a rigorous proof is still lacking, it should be
worthwhile to report on further progress. Based on the
random-cluster representation of the model,  we investigate the structure of 
clusters in terms of their dissociation probabilities under resampling of 
locations, and show that clusters with positive dissociation probabilities 
are unlikely to occur uniformly in the activity $z$ when $q$ is large.
(We believe that these ideas will eventually lead to a rigorous proof of 
first-order phase transition, but at this stage there are still a number of
difficulties to overcome.) On
the other hand, we have undertaken a numerical study of these problems. We 
use a natural continuum analog of the Swendsen--Wang cluster algorithm which 
was originally designed for the lattice Potts model. 
Arguing that a coexistence of ordered and disordered phases manifests
itself as a medium-term dependence
of the algorithm on the initial conditions, we find that for any $q$ there 
is only one critical activity $z_c$, and the phase transition at $z_c$ is 
of second order when $q = 2,3,4$, while it is of first order for $q\ge 5$.  

This picture confirms the main results of earlier numerical studies 
\cite{machta97,machta96,machta00} based on the so-called invaded
cluster algorithm.  However, the earlier results were not
conclusive about the order of the transition in the $q=4$ case which
we can better determine now. 
In addition, the exact relationship between the stationary measure of the 
invaded cluster algorithm and the corresponding finite
volume Gibbs measures of the Potts model is not known.  It has
only been postulated that in the infinite volume limit there should be
a unique activity $z$ such that they coincide, but this 
remains still to be proven.

Therefore, we felt it necessary  
to make an independent numerical study using an algorithm whose stationary
measure we can show to be a Gibbs measure of the Potts 
model.  This enables us to check, at least numerically, some of the
results obtained by the invaded cluster algorithm.
As already mentioned, our results confirm
qualitatively the earlier ones, and it is plausible that the invaded
cluster algorithm describes the continuum Potts model in the infinite
volume limit.  However, we also found a case in which
the invaded cluster algorithm gave results with significant 
finite size bias, slowly decreasing with the volume.  Such behavior makes
the control of 
finite size effects in the invaded cluster algorithm difficult,
and a careful analysis of the bias would be advisable before the results
are actually applied to describe the Gibbs states of the Potts model.

The outline of this paper is as follows: In Section~2 we introduce the
continuum Potts model and the continuum Swendsen--Wang algorithm.
In Section~3 we analyze the typical structure of clusters in the high-$q$
continuum Potts model in terms of their dissociation probabilities, 
and discuss the behavior  
of the algorithm in the presence of first-order phase transitions.
Section~4 provides our simulation results, while Section~5 contains a detailed 
discussion including a comparison with the results obtained by the invaded  
cluster algorithm, and a discussion on the structure of pure phases. 
 
\section{Model and algorithm} 
 
\subsection{Potts model in the continuum} 
 
The continuum Potts model is a model of point particles having $q \geq 2$ 
different types  and sitting in a rectangular box $\L\subset\R^d$, 
$d\ge2$. Rather than of particles  of different types, one may also think 
of particles with a ferromagnetic spin with $q$  possible orientations. 
A configuration of particles in $\L$ is given by a   
pair $\bX=(X,\s)$, where $X$ is the set of occupied 
positions, and $\s:X\to\{1,\ldots,q\}$ is a mapping attaching 
to each particle in $X$ its type, or ``color''. 
Writing $X_a=\{x\in X:\s(x)=a\}$ for the 
configuration of particles of type $a$, we may also think of $\bX$ as the 
$q$-tuple of the pairwise disjoint sets $X_a$ belonging to 
$\cX_\L=\{X\subset\L: \# X<\infty\}$,  
the set of all finite subsets of $\L$. The configuration space is thus 
equal to  $\cX_\L^{(q)}$, the set of $q$-tuples of pairwise disjoint 
elements of $\cX_\L$.  The particles are 
supposed to interact via a repulsive interspecies pair potential 
$\ph:\R^d\to[0,\infty]$ of bounded  support. For simplicity we confine 
ourselves to the case of a step potential 
\[ 
\ph(x{-}y) = \left\{\ba{cl}1&\text{if $|x{-}y|\le 1$,}\\ 
0&\text{otherwise}  \ea\right. 
\] 
already considered, e.g., in \cite{LL72,SH64}. The Hamiltonian in $\L$ is 
thus given by  
\be{Cont-Potts}  
H_\L(\bX)=\sum_{1\le a<b\le q} \ 
\sum_{x\in X_a,\, y\in X_b}\ph(x{-}y)\;.   
\ee  
Here we impose periodic boundary conditions, 
meaning that the difference $x{-}y$ has  to be understood modulo $\L$. (We 
note in passing that one could also add a molecular, type-independent 
interaction term, as was done in \cite{GH96}. Here, however, we stick to 
the simple case above.)   The associated Gibbs distribution 
with activity $z>0$ and  temperature $T\ge0$ is then  
\be{cont-Gibbs} 
\mu_{\L,z,T}(d\bX)= Z_{\L,z,T}^{-1} \;  
\exp[- \,H_\L(\bX)/T] \;\prod_{a=1}^q z^{\# X_a}\,L_\L(dX_a)\,,  
\ee  
where $L_\L$ is the Lebesgue-Poisson measure on $\cX_\L$ defined by 
\[ 
L_\L(A) = \sum_{n\ge0}\frac{1}{n!} 
\int_{\L^n}1_A(\{x_1,\ldots,x_n\})\;dx_1\cdots dx_n 
\] 
and $Z_{\L,z,T}$ is the normalizing constant. The zero-temperature case   
corresponds to the classical model of Widom and Rowlinson \cite{WR70} 
with hard-core interspecies repulsion. One can then imagine that 
the particles form balls of diameter $1$ which can overlap
only when they are of the same type.

The behavior of the infinite-volume Gibbs states of this model is 
completely understood when $z$ is either small or large. If $z$ is small 
enough,  there is a unique infinite-volume state which is disordered, in 
that it is invariant  under permutations of particle types; this can be 
seen, for example, by using disagreement percolation; cf. Proposition 7 of 
\cite{G00}. If $z$ is  sufficiently  large (depending on $T$), there exist 
$q$ distinct phases which are ordered, or demixed, in that one particle 
type is more frequent than  all other types; see \cite{CCK95,R71} for the 
case $T=0$ and \cite{GH96,LL72} for general $T$.  It is expected, but not 
rigorously known, that there exists a sharp activity threshold 
$z_c=z_c(T)$ such that the infinite-volume Gibbs state is unique when 
$z<z_c$  and non-unique for $z>z_c$. (This lack of knowledge is due to 
the fact that the model does not have any useful stochastic monotonicity 
properties. The only monotonicity known is that the particle density is an 
increasing function of $z$; cf. Section 4.2 of \cite{G00} and equation 
\eqref{eq:rhoprime}.) If $q$ is large enough, it is further 
expected that the transition at $z_c$ is of first order, meaning that the 
disordered and the $q$ ordered  phases exist simultaneously. 
This is the problem we address in this paper.
 
\subsection{Random-cluster representation} 
\label{sec:rcrep} 
 
Just as the lattice Potts model, the continuum Potts model admits a 
random-cluster representation of Fortuin-Kasteleyn type; see \cite{GS90} 
and the references therein, as well as \cite{CCK95,GH96}. This 
random-cluster representation will become important in the following. The 
random-cluster measure associated to \eqref{cont-Gibbs} is a probability 
measure for random graphs $\G=(X,E)$ in $\L$.  The vertex set $X$ is 
obtained from the configuration $\bX=(X,\s)$ by disregarding the particle  
types described by $\s$, and the edge set $E$ is obtained by drawing random  
edges between the points of $X$. Specifically, for each $X\in\cX_\L$ let  
$\cE_X$ consist of all sets of non-oriented edges between pairs of distinct  
points of $X$, and $\nu_{X,T}$ be the probability measure on $\cE_X$ for  
which an edge between a pair $\{x,y\}\subset X$ is drawn, independently  
of all other edges, with probability $p(x{-}y)=1-e^{-\ph(x{-}y)/T}$; as  
before, the difference $x{-}y$ is understood modulo $\L$.    
(In the Widom--Rowlinson case of hard-core interspecies repulsion, the  
randomness of the edges disappears in that all points of distance $\le1$  
are connected automatically.) 
 
The random-cluster measure associated to \eqref{cont-Gibbs} thus 
lives on the space 
$\cG_\L=\{\G=(X,E): X\in\cX_\L,\; E\in\cE_X\}$ of all finite graphs in $\L$, 
and is given by 
\be{cont-RC} 
\chi_{\L,z,T}(dX,dE) = Z_{\L,z,T}^{-1}\; z^{\#X }q^{k(X,E)}\,  
L_\L(dX)\,\nu_{X,T}(dE)\;, 
\ee 
where $k(X,E)$ stands for the number of clusters of the graph $(X,E)$, and 
$Z_{\L,z,T}$ again denotes the normalization constant. 
(Note that this definition makes sense for any real $q>0$.) 
As indicated by our notation, the normalization constant is in fact the 
same in either of equalities \eqref{cont-Gibbs} and \eqref{cont-RC} for any  
allowed values of the parameters. This was established in \cite{GH96} as  
part of the proof for the following precise relationship between the two  
measures: 
\begin{proposition}\label{prop:cont-RC}  
{\rm($\mu\car\chi$)} Take a particle configuration
$\bX=(X,\s)\in\cX_\L^{(q)}$ with  
distribution $\mu_{\L,z,T}$ and define a random  graph $(X,E)\in\cG_\L$ as 
follows: Independently for each pair $\{x,y\}$ of 
points of the same type (i.e., $\s(x)=\s(y)$) 
let $\{x,y\}\in E$ with  probability $p(x{-}y)=1-e^{-\ph(x{-}y)/T}$. Then 
$(X,E)$ has distribution $\chi_{\L,z,T}$. 
 
{\rm ($\chi\car\mu$)}  Pick a random graph $\G=(X,E)\in\cG_\L$ according to 
$\chi_{\L,z,T}$ and define a type assignment $\s$ as follows:  
For each cluster $C$ of $\G$ assign a type $a\in\{1,\ldots,q\}$  
independently and with equal probability, and then 
define $\s(x)=a$ for all $x$ in the union of all clusters of the type $a$. 
Then $\bX=(X,\s)$ has distribution $\mu_{\L,z,T}$. 
\end{proposition} 
 
To obtain a joint picture of the continuum Potts model and its 
random-cluster representation, one should think of cluster-colored graphs 
$\G^\s=(X,E,\s)$, where $\G=(X,E)\in\cG_\L$ and $\s$ is a mapping attaching 
to each cluster $C$ of $\G$ a color $\s(C)\in\{1,\ldots,q\}$.  Let us use 
the notation $\mean{f}_{\L,z,T}$ for expectation values of a  
random variable $f$ on the cluster-colored graphs.  The continuum 
Potts model is then obtained by interpreting $\s$ as a function on $X$ which 
is constant on all clusters, and then forgetting the 
edges $E$; that is, for any $f$ which depends only on $\bX=(X,\s)$, 
$\mean{f}_{\L,z,T} = \int\!\mu_{\L,z,T}(\rmd\bX)\, f(\bX)$.  
Likewise, forgetting the colors one arrives at the random-cluster 
measure: for $f$ depending only on $(X,E)$, we have  
$\mean{f}_{\L,z,T} = \int\!\chi_{\L,z,T}(\rmd X,\rmd E)\, f(X,E)$.  
 
\subsection{Conditional single-type distributions} 
 
The Gibbs distribution $ \mu_{\L,z,T}$ in \eqref{cont-Gibbs} has also 
another useful property easily to be exploited for simulation. Namely, if 
we fix all particles except those of a given type $a\in\{1,\ldots,q\}$, 
then the conditional distribution of the particles of type $a$ is Poisson 
with a simple intensity function.  Specifically, for any bounded 
non-negative function $u$ on $\L$ let 
\be{Poisson} 
\pi_\L^u(dX) =\exp\big[-{\textstyle\int\limits_\L u(x)\,dx}\big]  
\prod_{x\in X}u(x) \ L_\L(dX) 
\ee 
be the Poisson point process on $\cX_\L$ with intensity function $u$. Then 
the following observation follows immediately from the definitions 
\eqref{Cont-Potts} and \eqref{cont-Gibbs}: 
\begin{proposition}\label{prop:cond_single_type} 
Let $1\le a\le q$ be a given type and, for any $\bX=(X,\s)\in \cX_\L$,  
let $X_{\ne a}=\{x\in X:\s(x)\ne a\}$ be the set of positions of all 
particles having  
types different from $a$. Then, under $\mu_{\L,z,T}$, the conditional 
distribution of $X_a$ given $X_{\ne a}$ is equal to the Poisson point 
process $\pi_\L^{z\,p(\,\cdot\,|X_{\ne a})}$ with intensity function 
$z\,p(\,\cdot\,|X_{\ne a})$, where 
\[ 
p(x|X_{\ne a})=\exp\Big[-\sum_{y\in X_{\ne a}}\ph(x{-}y)/T\Big]\,. 
\] 
\end{proposition} 
 
To simulate $\pi_\L^{z\,p(\,\cdot\,|X_{\ne a})}$ one can use the well-known 
fact  
that $\pi_\L^{z\,p(\,\cdot\,|X_{\ne a})}$ can be obtained from the homogeneous 
Poisson point process $\pi_\L^z$ with constant intensity function $z$ by a 
random thinning: each point $x$ from a $\pi_\L^z$-sample $X$ is kept, 
independently of all other points, with probability $p(x|X_{\ne a})$; 
otherwise $x$ is removed. In the spirit of the random-cluster 
representation discussed above, this can also be achieved by independently 
drawing (virtual) edges between the points $x$ of $X$ and $y$ of $X_{\ne 
a}$ with probability $p(x{-}y)$, and deleting all $x\in X$ that are 
connected by an edge to some $y\in X_{\ne a}$. 
 
\subsection{The continuum Swendsen--Wang algorithm} 
 
The algorithm of Swendsen and Wang \cite{SW87} is by now a standard device 
for simulating the lattice Ising and Potts models. It can be characterized 
as the algorithm which alternatively applies the transition probabilities 
relating the Potts model with its random-cluster representation.  The naive 
analog for the continuum Potts model would be an alternative application of 
the two steps described in Proposition \ref{prop:cont-RC}. Note, however, 
that these steps always keep the set of occupied positions fixed. That is, 
these transition  steps are unable to equilibrate the particle 
positions. (Iterating these steps,  one would rather arrive at the discrete 
random-cluster distribution of edges between the vertices chosen 
initially.) So, one has to combine these steps with a further simulation 
step which takes care of the positions. The simplest such step is the Gibbs 
sampler based on the conditional probabilities of Proposition 
\ref{prop:cond_single_type}. We are thus led to the following continuum 
version of the Swendsen--Wang algorithm, variants of which have already been 
proposed independently in \cite{CM98} and \cite{HLM99}: 
 
\bigskip 
\noindent{\bf Continuum Swendsen--Wang algorithm:} 
Start from any initial configuration $\bX\in\cX_\L^{(q)}$ and iterate the 
sweep   
consisting of the following three steps: 
\begin{description} 
\item[\rm CSW 1:] \emph{Resampling of positions. }Successively for 
$a=1,\ldots,q$,  replace $X_a$ by a sample from the Poisson point process 
$\pi_\L^{z\,p(\cdot|X_{\ne a})}$, using a random thinning of $\pi_\L^{z}$. 
\item[\rm CSW 2:]  \emph{Drawing edges. }Let $X=\bigcup_{a=1}^q X_a$ and, 
independently for each pair $\{x,y\}$ of points of the same type, draw an 
edge from $x$ to $y$ with probability $p(x{-}y)=1-e^{-\ph(x{-}y)/T}$. Let 
$E$ be the resulting set of edges, and consider the graph $\G=(X,E)$. 
\item[\rm CSW 3:]  \emph{Choice of types. }For each cluster $C$ of $\G$, 
independently of all other clusters, pick  a random type uniformly in 
$\{1,\ldots,q\}$ and assign this type to each $x\in C$. Let $X_a$ be the 
set of vertices receiving type $a$, and $\bX=(X_1,\ldots,X_q)$. 
\end{description} 
In view of Propositions \ref{prop:cont-RC} and \ref{prop:cond_single_type}, 
it is clear that the Gibbs distribution $\mu_{\L,z,T}$ is invariant under 
this algorithm.  In fact, the following ergodic theorem holds: 
\begin{proposition}\label{prop:ergodicity} 
Let $\bX_n\in\cX_\L^{(q)}$, $n\ge 0$, be the realization of the continuum 
Swendsen-Wang algorithm after $n$ sweeps. Then the distribution of $\bX_n$ 
converges to  $\mu_{\L,z,T}$ in total variation norm at a geometric rate. 
\end{proposition} 
\begin{proof} It suffices to observe that, in Step CSW 1, $X_a=\emptyset$ 
for all $a$  
with probability at least $\d=e^{-zq\,|\L|}$. So, if $\bX_n$ and $\bX_n'$ 
are two versions of the process starting from different initial 
configurations but otherwise using the same realizations of randomness, then 
\[ 
\mbox{Prob}(\bX_n\ne \bX_n')\leq (1-\d)^n\;, 
\] 
whence the proposition follows immediately. 
\end{proof} 
For practical purposes, particularly in our context, the ergodic theorem   
\ref{prop:ergodicity} is rather misleading. This is because the rate of  
convergence towards $\mu_{\L,z,T}$ can be extremely small, even for $\L$  
of moderate size. (This is already seen from the number $\d$ above,  
though this is only a simple lower estimate of the coupling probability.) 
In particular, this is the case in the presence of a first-order phase 
transition when $\mu_{\L,z,T}$ is essentially supported on disjoint sets 
$A_i$, $i=0,\ldots,q$, that are typical for the $q+1$ coexisting phases. 
In this case, the sets $\Adis=A_0$ and $\Aord = \cup_{i=1}^q A_i$ are  
separated by tight bottlenecks of the CSW-algorithm. In fact, over a  
fairly long initial period the CSW-algorithm will converge to the  
conditional probability $\mu_{\L,z,T}(\cdot\,|\,A)$, with 
$A=\Adis$ or $\Aord$ depending on the initial condition,
and $\mu_{\L,z,T}$ is reached only after a time  
that is far exceeding any reasonable observation period. In this way, one  
can detect a first order phase transition by comparing the CSW-algorithm  
for different initial conditions. We will discuss this point in more detail  
in Subsection \ref{sec:absorption}. 
 
We conclude this section comparing the algorithm described above 
with the related algorithm invented in \cite{CM98,HLM99}. 
\begin{remark}\label{rem:CM} 
Instead of the systematic scan through all types in Step CSW 1 above, one 
could also use a random scan by resampling only the positions of a random 
(or, by type symmetry, a fixed) type. Contracting our Step CSW 3 with the 
successive Step CWS 1 (for a single $a$) one then arrives at the following 
algorithm proposed in \cite{CM98,HLM99}: 
 
\smallskip  
{\it Random-scan continuum Swendsen--Wang algorithm:} Starting 
from any initial graph $\G\in\cG_\L$, iterate the following two steps: 
\begin{enumerate} 
\item Each cluster $C$ of $\G$, independently of all others, is deleted 
with  probability $1/q$ and retained with probability $(q-1)/q$. Let 
$(X_{\text{old}}, E_{\text{old}})$ be the 
remaining graph. 
\item  Choose a sample $X_{\text{new}}$ from 
$\pi_\L^{z\,p(\cdot|X_{\text{old}})}$ and, independently for each 
pair $\{x,y\}$ in $X_{\text{new}}$, draw an edge from $x$ to 
$y$ with probability $p(x{-}y)$.  Let $E_{\text{new}}$ be the 
resulting set of edges, and consider the graph   
$\G=(X_{\text{old}}\cup X_{\text{new}},  
E_{\text{old}}\cup E_{\text{new}})$. 
\end{enumerate} 
While this version has the advantage of remaining completely in the 
random-cluster picture (and thus working also for non-integer $q$), a 
priori it is not  evident whether it is more efficient or not.  Clearly, 
$q$ sweeps of the random-scan version require the same numerical effort as 
one sweep of the systematic-scan version.  However, the former has higher 
correlations even after $q$ sweeps since the  waiting time until \emph{all} 
clusters are resampled (the maximum of the geometric waiting times for 
replacement of a single cluster) has expectation larger than $q$, and thus 
does  not seem to converge at a geometric rate not depending on the number 
of clusters.  
We performed a brief numerical comparison of the efficiencies and behavior 
of the above two algorithms, and these support the above picture: both 
algorithms lead to similar behavior but the 
systematic scan is slightly more efficient than the random scan. 
\end{remark} 
 
\section{Detecting first-order phase transitions in finite volume} 

\subsection{The structure of clusters for high $q$}
\label{sec:highq} 

In this subsection we present a rigorous result showing a common feature of
\emph{all}\/ clusters at \emph{any}\/ activity 
when $q$ becomes large, in arbitrary dimensions.   
We recall first what happens in the planar 
lattice Potts model; see \cite{KS82}.   
For large $q$ it is known that, with probability one, the plaquettes  
($2\times 2$ squares) on which the configuration shows one particular  
of several typical patterns form an infinite cluster. In the ordered  
phase with dominating spin value $a$ the pattern corresponds to the  
local ground state in which spins take the same value $a$, while in  
the disordered phase the pattern corresponds to a local ceiling state  
in which all nearest neighbor spins differ.  
Configurations which belong to neither category fail to get weight in  
the thermo\-dynamic limit. In the continuum Potts system we expect a  
similar characterization in terms of percolation: an ordered phase with  
dominating type $a$ should be characterized by percolation of spins of  
value $a$, while the disordered phase by percolation of vacancies.  
Moreover, it should be possible to derive these properties from certain  
typical local patterns characteristic of the ordered or disordered case.  
Corollary \ref{cor:scenarios} below will show which kind of local patterns 
can occur for any $z$ and any phase when $q$ is large. 

In the following we confine ourselves to the
Widom--Rowlinson case $T=0$. (We believe that similar estimates should
hold also for $T>0$, but this would need extra effort.)
Consider a configuration $X\in\cX_\L $ in a box $\Lambda$. Since we are 
in the Widom-Rowlinson case, the associated set of edges in the random  
cluster model is deterministic, viz.\
$E_X\equiv\big\{\{x,y\}\subset X: |x-y|\le1\big\}$.  
Let $C\subset X$ be a cluster of the graph $\Gamma=(X,E_X)$. We write  
\[ 
{\cal U}(X\setminus C) =\big\{x\in\Lambda: \exists \,y\in X\setminus C,\  
|y-x|\le 1\big\} 
\] 
for the part of $\L$ in which any point is connected to $X\setminus C$, 
and 
$ 
\D_C(X) = \Lambda \setminus {\cal U}(X \setminus C) 
$ 
for the available free space of $C$. We consider the probability 
\begin{equation} 
\d_C(X) = L_{\D_{C}(X)|\#C} (\xi: k(\xi) \geq 2 ). 
\end{equation} 
In the above, $ \#C$ is the number of particles of $C$,  
\[ 
L_{\D|N}(A) = |\D|^{-N} 
\int_{\D^N}1_A(\{x_1,\ldots,x_N\})\;dx_1\cdots dx_N 
\] 
is the distribution of a configuration of $N$ particles thrown independently 
and uniformly into $\Delta$ (where $|\D|$ is the Lebesgue measure of $\D$), 
and $k(\xi)$ is the number of clusters of $(\xi,E_\xi)$. 
We call $\delta_C(X)$ \emph{dissociation probability}, for it measures  
how big the chance is to split $C$ into two disconnected parts by a random  
resampling of its points in the room available after taking $C$ away. 
In particular, a small dissociation probability makes it unlikely
for $C$ to admit a pivotal point which cannot be removed without splitting
$C$ into disconnected parts, and therefore expresses some kind of robustness
of $C$. The following result states that, for large $q$, all clusters are
robust in this sense, and this is a property uniform in $z$. 
 
\begin{proposition}\label{prop:dissoc_prob} 
Let $T=0$ and $\kappa(\L)=\max\{k(X): X\in \cX_\L\}$ be the maximal  
number of clusters in $\L$. Then for all $q>0$, $\d>0$ and $z>0$ we have 
\[ 
\chi_{\L,z,0}\Big(X \in \cX_\L: \exists \text{ cluster }C\subset X,\  
\delta_{C}(X) \geq \delta \Big) \leq 2^{\kappa(\L)}/\delta q\,. 
\] 
 \end{proposition} 
Note that $\kappa(\L)$ is of the order $|\L|$ when $\L$ is a square. 
Therefore, this estimate cannot be applied directly to the infinite volume 
limit.  The main point of the bound is that it is uniform in the activity 
$z$, and thus applies to all possible finite-volume ``phases'' for 
sufficiently high values of $q$. We have not tried to optimize the 
constant in the Proposition, and $2^{\kappa(\L)}$ is most likely very 
far from being optimal. 

Postponing the proof for a moment, let us first dicuss the consequences  
of this result.  
Intuitively, a weak dissociation tendency of a cluster $C$ means that either 
$C$ is a singleton (in which case it cannot dissociate), or the available  
free space $\Delta_{C}(X)$ is small compared to the number of particles.  
One instance of the latter case is captured by the following definition: 
For any given $0<\g<1$, let a cluster $C$ in a
configuration $X$ be called \emph{$\g$-confined}
if $\# C\ge2$ and $\D_{C}(X)$ 
admits no two Borel subsets $\D_1$ and $\D_2$ such that they are
at least a unit distance apart and each contains
a fraction $\gamma/2$ of the total free volume, i.e., such that
$\mathrm{dist}(\D_1,\D_2)>1$ and $|\D_1|,|\D_2|\ge \g|\D_{C}(X)|/2$.
Of course, this condition means that $\D_{C}(X)$ must be small, 
in that it either has diameter at most one or consists of a solid core 
exceeding the diameter $1$ only by some tiny filaments (possibly scattered 
all through $\L$). 

\begin{corollary}\label{cor:scenarios} 
Let $T=0$, $0<\g<1$,  a large number $N_0 \in \mathbb{N}$, and $\e>0$ be  
given. Then there exists some $q_0 \ge 1$ such that, for all $q \ge q_0$  
and $z > 0$, 
 \[ 
\chi_{\L,z,0}(\Adis_\L\cup\Aord_\L)\ge 1-\e, 
\] 
where  
$\Aord_\L= \{X\in\cX_\L: \exists\text{ cluster }C\subset X,\ \#C\ge N_0\}$ and 
\[ 
\Adis_\L =\{X\in\cX_\L: \forall\text{ clusters }C\subset X,\ \#C=1 \text{ or }
C \text{ is $\g$-confined}\}\,.
\] 
 \end{corollary} 
 
Evidently, the event $\Adis_\L$ describes a scenario characteristic for the
disordered  phase: For very small $z$, all clusters will typically be
singletons, and the model is similar to the hard-core gas of balls of unit
diameter. If $z$ increases,  the singletons (considered as balls) will become
more and more densely packed.  At the close-packing density of balls the color
entropy cannot be increased any more, and the increasing particle density will
force the system to build up  confined clusters of two or more overlapping
balls. But all clusters are still separated
by channels of vacancies.  At the threshold $z_c$, color entropy breaks down
in favor of positional entropy, which means that the system will form a large
cluster $C$ of size $\# C\ge N_0$, where particles have more positional 
``degrees of freedom''.
It is then plausible that $\Delta_{C}(X)$  fills a macroscopic part of
$\L$ and all other clusters are confined.
Since the color is constant on $C$, we could conclude that a fixed spin value
$a$   is dominating with overwhelming probability. So, with a proper choice of
$N_0$, the $\Aord_\L$ scenario should be typical for the ordered phases above
$z_c$.
 
We now turn to the proofs of Proposition \ref{prop:dissoc_prob} and 
Corollary \ref{cor:scenarios}. 
 
\begin{proof}[ of Proposition \ref{prop:dissoc_prob}] 
We start by noting a symmetry property of the Lebesgue-Poisson measure $L_\L$ 
which readily follows from its definition: 
For any measurable function $F:\cX_\L^3\to [0,\infty[$, 
the expression 
\begin{equation}\label{eq:symmetry} 
\int L_\L(dX) \sum_{\xi\subset X} \int L_{\L|\#\xi}(d\eta) \, 
F(\xi,\eta, X\setminus\xi)  
\end{equation} 
is invariant under the exchange of the first two arguments of $F$. Now,  
\[ 
\begin{split} 
\d q \;& Z_{\L,z,0}\ \chi_{\L,z,0}\Big(X \in \cX_\L: \exists 
\text{ cluster }C\subset X,\   \delta_{C}(X) \geq \delta \Big)\\ 
&\le \int L_\L(dX)\ z^{\#X }\,q^{k(X)+1}  
\sum_{\xi\subset X} 1_{\{\xi\text{ cluster of }X\}}\, \d_{\xi}(X)\\ 
&\le \int L_\L(dX) \sum_{\xi\subset X} 1_{\{\xi\subset
  \D_{\xi}(X),\,k(\xi)=1\}}\,  
\\ 
&\qquad\times 
(|\L|/|\D_{\xi}(X)|)^{\#\xi}\int L_{\L|\#\xi}(d\eta)\,  
1_{\{\eta\subset\D_{\xi}(X),\,k(\eta)\ge2 \}}\,  
z^{\#(X\setminus\xi\cup\eta)} \,q^{k(X\setminus\xi\cup\eta)} 
\end{split} 
\] 
since $k(X)+1\le k(X\setminus\xi\cup\eta)$ under the circumstances described
by the indicator functions. 
Next we use the symmetry property of expressions of the form 
\eqref{eq:symmetry}, together with the fact that $\D_{\xi}(X)$ depends 
only on $X\setminus\xi$.  The last integral then becomes 
\[ 
\begin{split} 
&\int L_\L(dX) \sum_{\xi\subset X} \int L_{\L|\#\xi}(d\eta)\,  
1_{\{\eta\subset \D_{\xi}(X),\,k(\eta)=1\}}\, 
\\ 
&\qquad\qquad\times 
(|\L|/|\D_{\xi}(X)|)^{\#\eta}\,  
1_{\{\xi\subset\D_{\xi}(X),\,k(\xi)\ge2 \}}\, z^{\#X} \,q^{k(X)}\\ 
&=\int L_\L(dX) \, z^{\#X} \,q^{k(X)} \sum_{\xi\subset X} 
1_{\{\xi\subset\D_{\xi}(X),\,k(\xi)\ge2 \}} 
\  L_{\D_{\xi}(X)|\#\xi}\big(k(\cdot)=1\big)\,. 
\end{split} 
\] 
Finally,  we estimate away the last probability in the last integrand simply  
by 1 and note that the condition $\xi\subset\D_{\xi}(X)$ means that $\xi$ 
is disconnected from $X\setminus\xi$ and, therefore,
consists of a union of clusters of $X$. Since there  
are at most $2^{k(X)}\le 2^{\kappa(\L)}$ such unions of clusters, the last  
expression is not larger than $Z_{\L,z,T}\, 2^{\kappa(\L)}$, and the result  
follows. 
\end{proof} 
 
\begin{proof}[ of Corollary \ref{cor:scenarios}] Let $\d=\g^{N_0}/2$ and 
$q_0$ be so large that $2^{\kappa(\L)}/\delta q_0\le\e$.  
By Proposition \ref{prop:dissoc_prob}, it will be sufficient to show that 
$\d_C(X)\ge\d$ whenever $2\le\#C\le N_0$ and $C$ is not $\g$-confined. 
However, in this case there exist two Borel subsets $\D_1$ and $\D_2$  
such that $\mathrm{dist}(\D_1,\D_2)>1$ and $|\D_i|\ge \g|\D_{C}(X)|/2$ 
for $i=1,2$. So, a resampling of the points of $C$ within $\D_{C}(X)$ 
will certainly produce at least two clusters whenever all new points fall 
within $\D_1\cup\D_2$, and each of the sets $\D_1$ and $\D_2$ gets 
at least one of them. Hence 
\[ 
\d_C(X)\ge \g^{\# C} \,\Big[1- \sum_{i=1}^2 
\big(\textstyle\frac{|\D_i|}{|\D_1|+|\D_2|}\big)^{\# C} \Big]  \ge \d, 
\] 
and the proof is complete. 
\end{proof}

\subsection{Quasi-absorbing sets of the CSW-algorithm} 
\label{sec:absorption} 
 
We now take up the discussion begun after Proposition \ref{prop:ergodicity}
on the behavior of the CSW-algorithm in the presence of first-order  phase
transitions. We will argue that a first-order phase transition is
characterized by the appearance of two different sets which are nearly
absorbing  for the CSW-algorithm, so that its behavior over a reasonable
observation period strongly depends on the initial condition. The occurrence
of such a dependence on the initial condition is therefore an indication of a
first order phase transition.

Our reasoning consists of three parts: First we will ask how a first order
transition  will manifest itself in finite volume, then we will study the
influence of a quasi-absorbing  set on the medium-term behavior of the
CSW-algorithm, and finally we explain why a jump of the particle density
should imply a bottleneck in the CSW-algorithm.

\medskip 1. How can a first-order phase transition be observed in finite
volume? 
By definition, a first order transition is identified by  a discontinuity of
the first derivative of the infinite  volume pressure.  In our case, with $z$
as parameter,   this corresponds to a discontinuity of  the particle density
at the transition point $z_c$.  It is conjectured  that for the continuum
Potts model with sufficiently large $q$, the percolation threshold $z_c$
should drive such a first order transition.  In addition, at $z_c$ one expects
a coexistence of disordered and ordered phases,  in that there exist two
mutually singular type-invariant Gibbs measures, $\mu^{\text{dis}}_{z_c,T}$
and $\mu^{\text{ord}}_{z_c,T}$,   which can be obtained as limits of the
unique type-invariant measures for $z\uparrow z_c$ resp.\ $z\downarrow z_c$,
 and are the only extremal elements of the set of type-invariant Gibbs
measures.  In particular, the average particle density of
$\mu^{\text{dis}}_{z_c,T}$  should be different from that of
$\mu^{\text{ord}}_{z_c,T}$.  Since the measure $\mu_{\L,z_c,T}$ is also
type-invariant,   its infinite volume limit would then be a (presumably
non-trivial)  convex combination of these measures.   There are then two
disjoint sets of configurations $\Adis_\L$  and $\Aord_\L$, approaching the
disjoint supports of $\mu^{\text{dis}}_{z_c,T}$ and $\mu^{\text{ord}}_{z_c,T}$
in the infinite volume limit in the sense that  $\mu_{\L,z_c,T}(\Adis_\L\cup
\Aord_\L)\to 1$ as $\L\uparrow\R^d$,   while separately each set has a
probability strictly bounded away from  zero, and the conditional measures
$\mu_{\L,z_c,T}(\cdot\,|\,A^{\text{dis/ord}}_{\L})$ have  average densities
which stay a fixed value apart from each other.  Since in both cases the
variance of the density  tends to zero in the infinite volume limit, we can
also assume the sets to  be chosen so that the density distributions are
are almost mutually singular.

\medskip 2. What is the behavior of a Markov chain admitting a unique
invariant  measure $\mu$ almost concentrated on two disjoint sets $A_1$ and
$A_2$ such that the conditional measures $\mu(\cdot\,|\,A_i)$ can be
distinguished by an observable $f$? It is then plausible to expect that the
sets $A_i$ are nearly absorbing, in that the Markov chain stays within these
sets with probability close to 1. The following remark explores the
medium-term behavior  of $f$ for such a Markov chain.
 
\begin{remark}\label{rem:first_order} Let $(\bX_n)_{n\ge 0}$ be a Markov 
  chain with a Polish state space $E$   and stationary distribution $\mu$, $A$
a measurable subset of $E$ with  $\mu(A)>0$, $\mu_A=\mu(\,\cdot\,| A)$ the
associated conditional distribution  and $(\bX_n^A)_{n\ge 0}$ the induced
Markov chain on $A$ with invariant  distribution $\mu_A$, and suppose
$(\bX_n^A)_{n\ge 0}$ is uniformly  ergodic.  Also, let $f:E\to\R^d$ be any
measurable observable (not constant  on $A$)  and $\e>0$.  By the large
deviation principle for Markov chains  there exists then some $\d>0$ such that
\[ 
\mbox{Prob}\bigg(\Big|\frac1 N\sum_{n=1}^N f(\bX_n^A) - \int f\,
d\mu_A\Big|\ge \e  \bigg) \le e^{-\d N}
\] 
for all $N\ge1$; see Theorems 6.3.8 and 2.3.6 of \cite{DZ}. Finally, suppose
that $A$   is nearly absorbing for $(\bX_n)_{n\ge 0}$, in that
\[ 
\mbox{Prob}(\bX_{n+1}\in A\mid \bX_n)\ge \g\quad \text{when }\ \bX_n\in A\,,
\] 
where $\g$ is so close to $1$ that   $\frac{-\ln
\a}\d<\frac{\ln(1-\a)}{\ln\g}$  for some prescribed error probability $\a$. It
then follows   that, for large but not too large $N$, the time average of $f$
along the first  $N$ steps of the original Markov chain $(\bX_n)_{n\ge 0}$ is
close to the   conditional expectation $ \int f\,d\mu_A$, with probability
close to $1$.  Namely,
\[ 
\mbox{Prob}\bigg(\Big|\frac1 N\sum_{n=1}^N f(\bX_n) - \int f\,  
d\mu_A\Big|\ge \e  \bigg) \le 2\a \quad \text{when } \bX_0\in A  \text{ and  }
\textstyle \frac{-\ln\a}{\d} \le N \le \frac{\ln(1-\a)}{\ln \g}\,.
\] 
Indeed, $(\bX_n)_{n\ge 0}$ coincides with $(\bX_n^A)_{n\ge 0}$ up to the
first time $T$ such that $(\bX_n)_{n\ge 0}$ leaves $A$, and $T$ is
geometrically distributed with parameter $1-\g$. Splitting the event in
question into two parts according to whether $T\le N$ or not we therefore
obtain the upper bound $1-\g^N+ e^{-\d N}$, and the result follows.
\end{remark} 
 
In other words, if a MCMC algorithm has a very small probability of leaving  a
set $A$, it typically stays for the whole observation period within $A$  and
thus, during this period, coincides with the induced Markov chain on  $A$
which converges to $\mu_A$. In the case under consideration, it turns  out
that the mixing properties of the induced Markov chains are good enough  for
this scheme to work for both a set $A=\Aord_\L$ of ordered configurations,
and a set $A=\Adis_\L$ of disordered configurations.  We cannot yet give a
proof for the existence of such absorbing sets  for the values of $q$ we
inspected here, but we  believe they have a characterization similar to what
was  given for the high-$q$ limit in Corollary \ref{cor:scenarios} (whence we
have chosen the same notation).  In fact, the simulations described below
support the existence of such absorbing sets and make clear that a first-order
transition of the continuum Potts model does indeed manifest itself by a
pronounced ``bottleneck'' of the CSW-algorithm between two different
quasi-absorbing sets.
 
\medskip 3. Finally we ask: Are there any indications that the CSW-algorithm
does indeed have a bottleneck between two different quasi-absorbing sets?  We
will argue that, for any given $\epsilon>0$, a jump greater than $\epsilon$ in
the particle density
in one sweep of the CSW-algorithm has probability tending to zero in the
infinite  volume limit (apart from the very first sweeps).  Therefore, if the
sets $\Aord_\Lambda$ and $\Adis_\Lambda$ are defined  in a way allowing a
distinction by different particle densities then,  near a first order
transition point and for large enough volumes,   these sets should be nearly
absorbing, and  passages from one to the other should create a bottleneck
after a large enough  number of sweeps, as this would require to pass through
a gap in the   density distribution of $\mu_{\L,z_c,T}$.
 
The density of a configuration can only be changed in step CSW 1.  Suppose
that the input configuration is $(X,\sigma)\in\cX_\L^{(q)}$.  We remove all
particles of type $a$, and consider the available  free volume, denoted by
$\D_{X_a}(X)$ as in Sec.~\ref{sec:highq}.  For the Widom--Rowlinson model, the
new particles of type $a$ are  distributed exactly according to a Poisson
measure with activity $z$  in this volume.  Since the density distribution
given by   the Poisson measure on a set with Lebesgue measure $v$   has
expectation value $z$ and variance $z/v$, the new set has a density $z$  with
fluctuations of the order of $1/\sqrt{v}$.  If $v\ll |\L|$, these  changes to
the total particle density are negligible, and if $v\gg 1$,   then the change
in the total particle density is sharply concentrated to  $(z v-\#X_a)/|\L|$.
When $T>0$, particles can also be added to the complement of  $\D_{X_a}(X)$,
but even then the probability is  significant only where the old particles are
not too dense, that is,  typically only near the boundary of the set
$\D_{X_a}(X)$.
 
If this is not the first sweep, then $X_a$ was obtained by  the same procedure
and should consist of regions which either are  small containing not too many
particles, or have particle density close to  $z$; see also the discussion in
Sec.~\ref{sec:highq} for the high-$q$ limit.     Then the above argument
indicates that typically  the particle density should fluctuate like a Poisson
density, that is   $\order{\smash{|\L|^{-1/2}}}$, in one sweep.  Our numerical
results, to be  discussed in the next section, agree with this scaling
relation,  apart from second order phase transition points.
 
\section{Numerical results} 

\subsection{Implementation of the algorithm} 
 
In this section we present our results from applying the CSW algorithm to
the {\em two-dimensional\/} Potts model.
The main numerical hurdle to overcome in the simulation of the algorithm  is
the large number of particles in the kind of volumes we need in order to  get
reliable estimates about the properties of the infinite volume phases.  Here
the bottleneck is not the memory required to store the  configurations, but
rather the time required to find those ``old'' particles  which lie within the
interaction radius  from a given ``new'' particle, as well as the time needed
for dividing  the particles into clusters according to a given edge
configuration.
 
The second part, the forming of clusters for a given configuration  of
particles and edges, can be done very efficiently by using the algorithm
described in \cite{newman01}.   We used the tree-based union/find algorithm
given in section II B.\ of \cite{newman01}, while simultaneously  keeping
track of the size  and  of the ``corners'' of the clusters (see section
\ref{sec:observables}  for the precise definition).
 
To overcome the large volume problem in the first part of the algorithm, we
used ``hashing'' of the box into smaller cells, altogether $\Nhash^2$ of
them.  The number of cells in one dimension,  $\Nhash$, was chosen so that
the average number of particles in the cell,  as determined by the Poisson
process with activity $z$, would be about $10$.  For each cell, we created
$q$ directed lists, one for each possible type.  The list  number $a$
contained pointers to all those particles which had the type $a$ and which
were within the interaction radius (here the unit distance)  from the cell.
 
This information was used to substantially reduce the time needed both  in
creation of the thinned Poisson configuration, and in computation of the  open
bonds: by construction,   if we add a particle of type $a$ anywhere  in the
cell, then it can interact only with particles in one of  the lists of the
cell with $a'\ne a$.  The use of $q$ separate lists  allowed for easy removal
of the particles with a certain type which was  needed in the first part of
the algorithm.
 
All simulations were performed using square boxes of linear size $L$,  i.e.,
$\L = [0,L]^2$, with  periodic boundary conditions.  We employed three
different initial  conditions: a Poisson sample with activity $z$, either
assigning all  particles the type $1$, or choosing the types randomly,  and a
``disordered crystal'' where the particles lie in a certain dense square
lattice with alternating types.  The initial conditions with a uniformly
colored Poisson sample are called here ``ordered''.  The other two
alternatives -- randomly colored Poisson and the disordered crystal -- led  to
the same behavior (i.e., to measurements within error bars of each other) in
all those cases where we tried both.  Therefore, we call them collectively
``disordered'' initial conditions in the following.
  
\subsection{Measurements} 
\label{sec:observables} 
 
For measurement of the properties of the infinite volume Gibbs states, we 
employed the numerical CSW-update algorithm with several values of 
the box size $L$ to an initial configuration $\bX_0\in\cX_\L^{(q)}$.  After a 
preset number of steps, $n_{0}\ge 1$  
(called ``burn-in'' or equilibration period), 
we measured the value of an observable $f(\Gamma^\sigma)$ for the following 
consecutive  
$\nm>0$ steps, and obtained a sequence of samples $f_n = 
f(\G^\sigma_{n_0+n})$ for  
$n=1,\ldots,\nm$.  The values $n_0$ and $\nm$ were assumed to be chosen so 
large that the average would well approximate the corresponding expectation 
value, 
\[ 
 \frac{1}{\nm} \sum_{n=1}^{\nm} f_n  \approx  
 \mean{f}_{\L,z,T}. 
\] 
Finding a good choice for $n_0$ and $\nm$ was not straightforward.  It was 
particularly difficult near the phase transition points where 
we found out that, even for 
this cluster algorithm, the equilibration and  decorrelation times for 
relevant observables can be very large.  We chose, quite arbitrarily, $n_0 
=250$ and  $\nm=2500$ as a first guess, and increased these values when 
necessary; when quoting the results we will use the short-hand phrase ``using 
$n_0+\nm$ sweeps'' to give the actual values used in computation of the 
results. 
 
To estimate the error arising from the finiteness of $\nm$, we computed 
the standard deviation from new samples obtained by 
dividing the data into 
$10$ blocks: the block averages are less correlated than two consecutive 
samples, and  as long as $\nm>10\times \text{(decorrelation time)}$ this 
should yield a fairly reliable estimate of the error.  Explicitly, the 
``errors'' given later were obtained by defining, for 
$n_{\text{B}}=\nm/10$ and $k=1,\ldots,10$, 
\[ 
  \bar f_k = \frac{1}{n_{\text{B}}} \sum_{n=1}^{n_{\text{B}}}   
  f_{(k-1) n_{\text{B}} + n} , 
\] 
then computing the sample variance 
\[ 
  S^2_f = \frac{1}{9} \Bigl[ \sum_{k=1}^{10} (\bar f_k)^2 
 - \frac{1}{10} \Bigl(\sum_{k=1}^{10} \bar f_k\Bigr)^2\, \Bigr] 
 = \frac{1}{9}  \sum_{k=1}^{10} \Bigl(\bar f_k 
 - \textstyle \frac{1}{10} \sum\limits_{i=1}^{10} \bar f_i\Bigr)^2 
\] 
of $(\bar f_k)$, and estimating 
the standard deviation of the average of $(f_n)$ by $S_f/\sqrt{10}$. 
Some of the decorrelation times for large boxes were indeed very long (see, 
for instance, Figure \ref{fig:swq4}) and these more elaborate methods were 
required to get a sensible error estimate. 
 
For any given colored cluster configuration $\G^\s=(X,E,\s)$ 
we considered the following four observables $\rho$, $\rho'$, $\gamma$ 
and $\dperc$: 
 
\emph{1. Particle density  } 
\[ 
  \rho(\G^\s) =  {N(X)}/{L^2}\,, 
\]  
where $N(X)=\#X$ is the total number of particles.   
 
\emph{2. Slope estimator $\rho'$. }  
Since 
\begin{equation}\label{eq:rhoprime} 
  \frac{\partial}{\partial z} \mean{\rho}_{\L,z,T} 
 = \frac{1}{z}\mean{ N\rho}_{\L,z,T} - 
\frac{1}{z}\mean{N}_{\L,z,T} \mean{\rho}_{\L,z,T} 
  = \frac{L^2}{z} \text{Var}(\rho), 
\end{equation} 
the scaled sample variance of $\rho$, 
\[ 
  \rho'=  \frac{L^2}{z} \,  
 \frac{1}{\nm-1} \Bigl[ \sum_{k=1}^{\nm} \rho_k^2 
 - \frac{1}{\nm} \Bigl(\sum_{k=1}^{\nm} \rho_k\Bigr)^2\, \Bigr] 
\] 
estimates the derivative of $\mean{\rho}_{\L,z,T}$ with respect to $z$.  
This is not an observable in the previous sense,  
so we computed its error estimate by using as samples 
the 10 sample variances computed from the 10 sample blocks of $\rho$. 
Note also that if $\rho'$ remains bounded when $L\to\infty$, then  
the corresponding standard deviation of density in one CSW-step is 
$\order{1/L}$ near stationarity. 
 
\emph{3. Largest cluster size $\gamma$. } 
This observable measures  
the ratio of particles in the largest cluster: 
\[ 
 \gamma(\Gamma^\s) =   {\max_C \,(\#C) }/{N(X)}\,, 
\] 
where the maximum is taken over all clusters $C$ of 
$\G=(X,E)$.   
 
\emph{4. Percolation radius $\dperc$. } 
This quantity measures the spread-out of the clusters 
from a given $L$-independent set $S_0$ in the middle of $\L$. 
As reference set we used 
\[ 
  S_0 = \defset{x\in\L}{\exists y\text{ such that }|x-y|\le 1/2 
    \text{ and } \norm{y-(L/2,L/2)}_\infty \le 3/2}, 
\] 
that is, a central $4\times 4$ square with rounded corners. 
For $i=1,2$, let $b^{-}_i(\G)$ and $b^{+}_i(\G)$ denote the 
minimum and maximum of the coordinates of clusters 
with particles in $S_0$, i.e., 
\[ 
  b^{-}_i(\G) = \min_{C:C\cap S_0\ne\emptyset} \min_{x\in C} x_i, \qquad 
  b^{+}_i(\G) = \max_{C:C\cap S_0\ne\emptyset} \max_{x\in C} x_i, 
\] 
where $C$ runs through the set of all clusters of $\G$. 
Our definition for the percolation radius then reads 
\[ 
 \dperc(\G^\s) = \max_{i=1,2}\,\Bigl\{\frac{L}{2} - b^{-}_i(\G)-1, \, 
   b^{+}_i(\G)-\frac{L}{2}-1,0\Bigr\}. 
\] 
(This particular choice is adapted to the  
Widom--Rowlinson case, $T=0$,  
where the most natural percolating objects are the 
discs of radius 1/2, as explained in Sec.~\ref{sec:rcrep}. 
For convenience, we retained this definition also in the case of 
positive temperatures where it can appear to be unnecessarily complicated.)   
For all practical 
purposes, it is safe to think of $\dperc$ as the maximal distance 
the clusters of $\G$ percolate away from the central $4\times 4$ square. 
 
After determining the critical values of $z$, 
we also repeated some of the simulations near these values 
in order to find out how the particles are distributed between 
clusters of different size.  To this end, we built histograms for 
cluster sizes by using the observables 
\[ 
  \frac{1}{N(X)} \sum_{x\in X} 
  1_{\bigl\{\frac{\#C(x)}{N(X)}\in\Delta\bigr\}} 
 =  \sum_{C} \frac{\#C}{N(X)}  
 1_{\bigl\{\frac{\#C}{N(X)}\in\Delta\bigr\}} 
\] 
where $C(x)$ denotes that cluster of $\G$ which contains the particle $x$, 
and the second sum goes over all clusters of $\G$. 
They describe the portion of the particles in clusters with size  
(relative to $N$) in the interval $\Delta$.  Here the intervals were 
chosen by dividing $[0,1]$ into $100$ pieces, i.e., using  
$\Delta_k=[k-1,k)/100$, for $k=1,\ldots,99$, and  
$\Delta_{100} = [99/100,1]$.  We also measured the  
average ratio of particles in very small clusters, with sizes from $1$ to  
$100$.  
 
\subsection{Computation of the critical activity} 
 
For the computation of the critical activity for a fixed temperature $T$ we 
started from a box with side length $L=8$ and computed the above 
observables for several values of $z$ by using an equidistant grid in a 
suitable range, always for {\em both\/} disordered and ordered initial 
conditions.  
This allowed an inspection of the effect of initial conditions and, apart 
from a neighborhood of a first order transition, these values always agreed 
within the computed error bars. 
 
\begin{figure}  
  \begin{center} 
    { 
      \myfigure{width=0.9\textwidth}{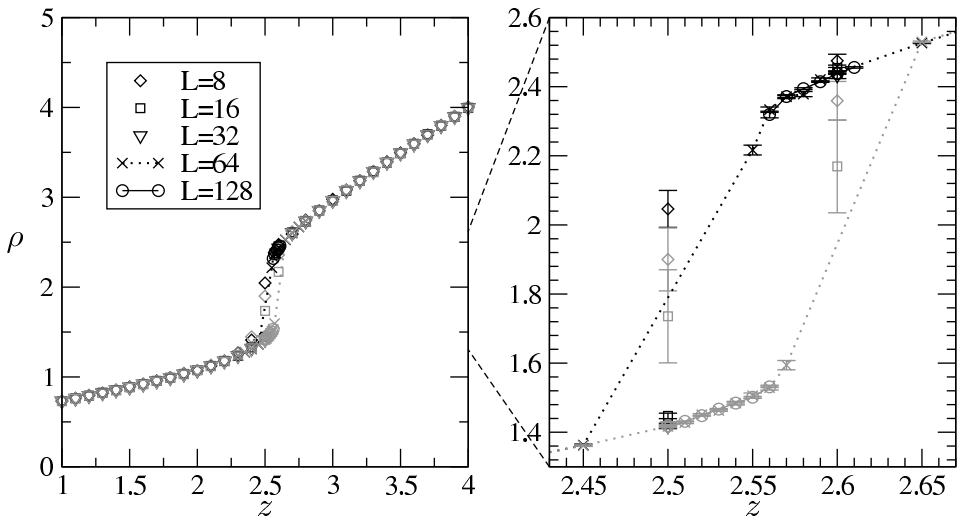} } 
    \caption{The measured values of $\rho$ as a function of $z$ for  
      $q=10$ and $T=0$ using five different box sizes and $250 + 2500$ 
      sweeps, all with both ordered (black) and disordered (grey)  
      initial conditions.  Only the equilibrated results have been 
      shown, see the text for explanation.} 
    \label{fig:q10rho} 
  \end{center} 
\end{figure}

It was expected that some percolation property could be an order parameter   
for the transition, and it turned out that both $\dperc$ and 
$\gamma$ had a pronounced change at the transition. Using the 
values measured for $\rho$, $\rho'$, $\gamma$ and $\dperc$ we could locate 
the critical value $z_c$ approximately.  The simulations were then 
repeated in a neighborhood of this value on a finer grid, but with twice 
the length $L$.  This was repeated until sufficient accuracy was achieved, 
typically at $L=128$, although we had to go up to box sizes $L=512$ for 
$q=4$ and $q=5$.  Figure \ref{fig:q10rho} shows the results of such an 
iteration for the density in the case $q=10$, $T=0$. 
 
\begin{figure} 
  \begin{center} 
    { 
      \myfigure{width=17em}{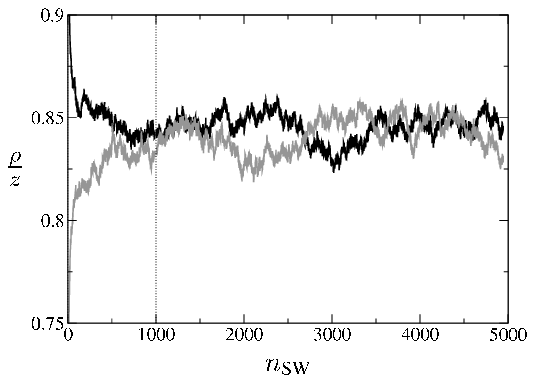} } \hfill 
    { 
       \myfigure{width=17em}{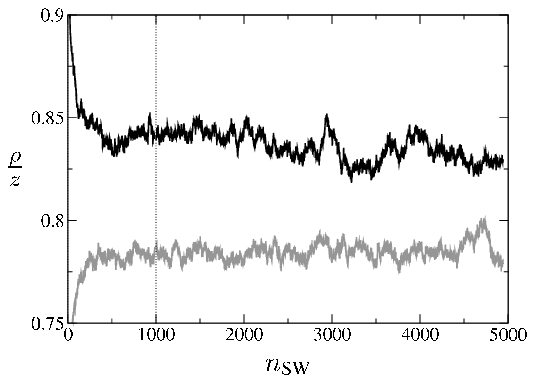}} 
     \caption{The evolution of $\rho/z$ under the algorithm 
       for $q=4$ (the left figure) and $q=5$ (the right figure),  
       and for two different initial conditions as described in the text:  
       ordered (the black line) and disordered (the grey line).  
       In all of these runs, $T=0$, $L=512$, and $z\approx \zcrit$ 
       ($z=2.05125$ for $q=4$ and $z=2.168$ for $q=5$). 
       $\Nsw $ denotes the number of ``sweeps'' performed, and   
       the dotted line represents the chosen equilibration cutoff, $n_0$.} 
     \label{fig:equilibration} 
  \end{center} 
\end{figure}

The order of the transition was determined from the dependence of the 
observables on using either ordered or disordered initial conditions.   If 
the different initial conditions led to different values of density, the 
transition was determined to be of first order.   
Our results also fully support the discussion made in
section \ref{sec:absorption} which allows us to identify the two different 
results as properties of the different coexisting phases. 
Figures \ref{fig:q10rho} and the $q=5$ part of \ref{fig:equilibration} 
present typical examples of the behavior in  
these instances:  both initial conditions lead to density fluctuations 
$\order{1/L}$, but the average values are separated by a constant 
$\order{1}$.  (This is true only for large enough $L$. For very small $L$, 
there is a significant probability to jump from one density region  
to another, 
and then the result is some average of the values for each phase, and the 
standard deviation is of the order of the gap between these values.) 
In addition, the observed values are right-continuous for the 
ordered initial conditions, and left-continuous for the disordered ones. 
 
In case the values agreed, 
we next looked at the behavior of $\rho'$: since its 
maximum was then always found to diverge as some power of $L$ in a 
neighborhood of the percolation threshold, we call these second 
order transitions.  In Figure 
\ref{fig:equilibration} we have plotted the evolution of the density under 
our algorithm for the values in the borderline cases: the largest $q$ 
for which the transition was found to be of second order, $q=4$, and the 
smallest $q$ for which it was of first order, $q=5$.  
 
\begin{figure} 
  \begin{center} 
    { 
      \myfigure{width=0.6\textwidth}{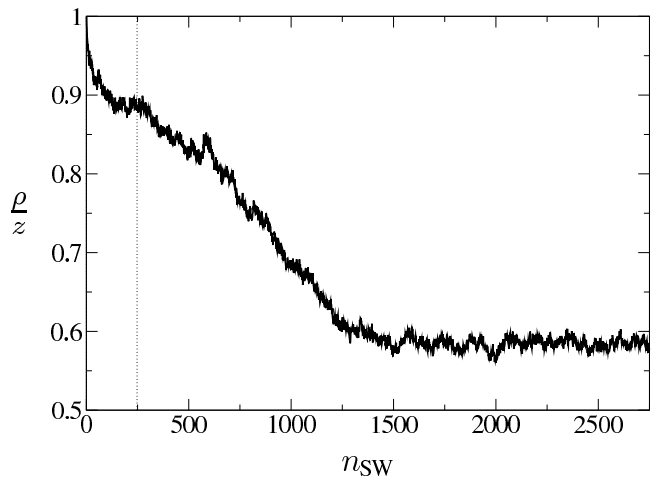} } 
     \caption{The evolution of $\rho/z$ for  
       $q=10$, $T=0$, and $z=2.54$, using $L=128$  and ordered initial  
       conditions. 
       $\Nsw $ denotes the number of sweeps performed, and  
       the dotted line the preset equilibration cutoff.} 
    \label{fig:q10therm} 
  \end{center} 
\end{figure}

In some cases, especially when using the ``wrong'' initial conditions 
near the borders of a first order transition region, the equilibration 
times turned out to be much longer than the chosen $n_0$ -- see  
Figure~\ref{fig:q10therm} for a typical instance.   
Similar problems,  
combined with very long decorrelation times, plagued the $q=4$ simulations  
in large boxes as well; note, for instance, that $n_0 = 250$ is clearly 
insufficient for  
the left part in Figure~\ref{fig:equilibration}.  However, since typically 
only one of the initial conditions suffered from these long equilibration 
times, we decided, instead of redoing the simulations with very large $n_0$, 
to throw away the non-equilibrated value and use only the equilibrated one.   
This explains some of the apparently ``missing points'' in 
Figure~\ref{fig:q10rho}.  Nevertheless, we {\em always\/} kept at least one 
result for each  
$z$, redoing the simulations with larger $n_0$ when necessary. 
 
\begin{table} 
\begin{center} 
\begin{tabular}{r|r@{.}lc|r@{.}lc} 
& \multicolumn{3}{c|}{$T=0$} & \multicolumn{3}{c}{$T=0.5$} \\ \cline{2-7} 
 $q$ & \multicolumn{2}{c}{$\zcrit$} & order  
& \multicolumn{2}{c}{$\zcrit$} & order \\ \hline 
 2 & 1 & 718(7) & 2nd & 1&86(4) & 2nd  \\ 
 3 & 1 & 907(8) & 2nd & \multicolumn{2}{c}{--} & -- \\  
 4 & 2 & 051(4) & 2nd & 2&273(7) & 2nd  \\ 
 5 & 2& 1675(13) & 1st & 2&424(2) & 1st  \\ 
 10 & 2&56(1) & 1st & 2&965(25) & 1st  \\ 
 50 & 3&65(30) & 1st & 4&95(65) & 1st  \\ \hline 
\end{tabular} 
\end{center} 
\caption{Estimates for the critical activity $\zcrit $  
and the order of the transition in the $L\to\infty$ limit for several $q$ 
and at zero and one non-zero temperature.  The error estimates  
are fairly conservative, see the text for how they were 
obtained from the simulations.}\label{t:criticalz} 
\end{table} 
 
\begin{figure} 
  \begin{center} 
    { 
      \myfigure{width=0.8\textwidth}{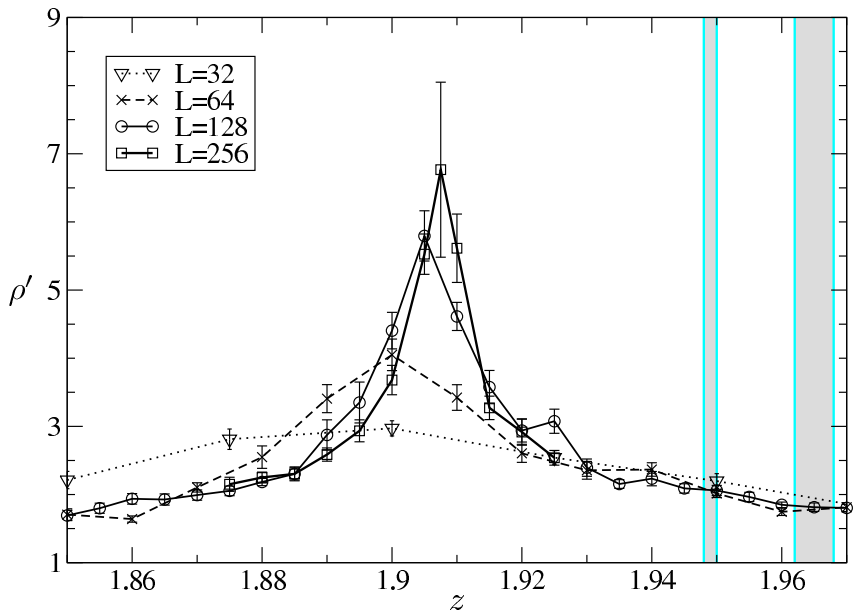} } 
    \caption{The measured values of $\rho'$ as the function of $z$ for 
      four different box sizes, computed for $q=3$ and $T=0$ 
      using $250+2500$ 
      sweeps.  The value shown is an average of the results 
      with ordered and disordered initial conditions. 
      The shaded regions depict the results given in \cite{machta97}  
      for the critical $z$ with $L=160$ (left region) and with $L=40$ (right 
      region).} 
    \label{fig:q3drho} 
  \end{center} 
\end{figure}

The results from this analysis are given in Table \ref{t:criticalz}.  For 
estimating finite-size effects, i.e., the difference of  
critical values from the $L\to\infty$ limit, we used certain  
monotonicity properties observed while doing the simulations.  For second 
order transitions, the observable $\rho'$ exhibits a 
divergence near the transition points, and the value quoted is the position of 
the maximum of the peak for the largest box size used.  
As can be seen also in Figure  
\ref{fig:q3drho}, this appears to increase monotonously with $L$, and 
therefore it can reasonably be taken as a lower bound for the 
limiting value.  The error given, $\delta z_c$, is a value such that  
for $z\ge z_c + \delta z_c$ the measurements of $\rho'$ for  
the two largest boxes agree with each other.  Again, as seen in  Figure  
\ref{fig:q3drho}, this is quite robust a value, fairly independent 
of $L$, and would in these cases be better understood as an upper bound for 
the error.  
For first order transitions, we used the property that the 
``coexistence window'' between the disordered and ordered phase goes to zero 
as $L\to\infty$, apparently monotonously.   
In these cases, the error is given by the range of values 
in which both initial conditions were equilibrated but yielded differing 
results, plus one grid spacing.  For instance, in the 
table we have $z_c = 2.56(1)$ for $q=10$, $T=0$, which was obtained from the 
single coexistence value shown in Figure \ref{fig:q10rho}. 
 
\section{Discussion} 
 
\subsection{Comparison with earlier results} 
 
The Widom--Rowlinson and continuum Potts models have already been studied 
numerically before in \cite{machta97} and \cite{machta00}.  These 
simulations used the so-called ``invaded cluster'' (IC) algorithm  
introduced in \cite{machta96} for studying the lattice Potts models near  
criticality.  This algorithm is similar to the Swendsen--Wang type of  
algorithms presented here. The main difference is that instead of  
generating samples for the finite volume Gibbs measure directly, a similar  
updating algorithm with a suitably chosen ``stopping rule'' is used for  
generating samples for a measure which differs from any of the finite  
volume Gibbs measures, but which is claimed to approach the correct Gibbs  
measure in the infinite volume limit. 
 
The main advantage of the IC algorithm is that an advance scanning of the  
parameter space for finding the transition values is not necessary, as the  
stopping rule is assumed to be chosen so as to force the system to be at  
the transition point in the infinite volume limit. Two stopping rules are  
used in \cite{machta97,machta00} to study the continuum models: the  
percolation rule for $q\le 3$ when the transition was expected to be of 
second order, and the fixed density rule for $q\ge 3$ to study whether the  
transition is of the first order. The main disadvantage of the IC algorithm  
is that the above mentioned convergence to the correct limit measure has  
not been proved so far and, in any case, it is quite difficult to control  
the finite size effects.   
 
In Figure \ref{fig:q3drho}, 
we have compared the results for the critical activity $z_c$  
for $q=3$ and $T=0$ from the present algorithm  
to those obtained using  
the percolating IC algorithm in \cite{machta97}. 
The finite size effects appear to be 
more prominent in the percolating IC algorithm: by Figure \ref{fig:q3drho} 
increasing the linear size fourfold from 
$L=40$ to $L=160$ does not appear even to halve the systematic error  
to the $L\to\infty$ value which we estimated (using lattice sizes up to 
$L=256$) from the position of the peak in $\rho'$. 
 
For $q=2$, $T=0$,  we found for the limiting value $z_c = 1.718(7)$,  while in
\cite{machta97} $z_c=1.7262(4)$ for $L=160$,
and $z_c=1.7201(7)$ for $L=40$ were measured.    The finite size effects seem
to be weaker in  this case.  In \cite{machta00}, simulations were performed
also   for a few non-zero temperatures.  Unfortunately, there is only one
instance in  which we can directly compare our results with theirs: for
$T=0.5$, $q=2$ we  estimated $z_c(L=\infty)= 1.86(4)$ while in \cite{machta00}
for $L=20$ with  the same   parameter values $z_c=1.8508(4)$ is given.  Due to
coarseness of our result, we  cannot really compare the finite size effects in
this case.
 
Let us offer a possible explanation for the  observed differences between the
results from these two methods.  Since percolation (i.e., the existence of a
cluster spanning the whole of $\L$)  was used as the stopping condition in the
above quoted  results, every sample configuration contains a percolating
cluster.    Even if we assume  that samples approximating the infinite volume
measure at the percolation  threshold are generated by the IC-algorithm, this
sampling method introduces  some bias into the measurements. Actually,
comparing the IC results with ours,  it appears that the IC-method
overestimates the critical activity $z_c$.   This can be understood by
observing that,  in finite volume, the probability   of having a percolating
cluster  is a continuous function of $z$ ascending from $0$ to $1$ around
$z_c$,  and thus is certainly not close to $1$ at $z_c$ but only when $z$ is
sufficiently larger than $z_c$.
 
For first order transitions, we would expect exactly the opposite  to happen:
the percolating IC-algorithm should underestimate $z_c$.    Indeed, since this
algorithm  produces only samples with a percolating cluster, the ordered phase
gains an advantage over the disordered phase whenever there is a chance for
it to occur. But, in finite volume, the ordered and disordered phases can
coexist throughout a whole range of parameter values $z$ (rather than only
\emph{at} $z_c$, as in infinite volume). The threshold detected by the
IC-algorithm therefore identifies only the lower end of the coexistence
interval. Unfortunately, we cannot test this hypothesis, as the fixed-density
stopping rule, and not the percolation one, was used in \cite{machta00} for
obtaining the critical activity for those   $q$ exhibiting first order
transitions.
 
Apart from these differences, our results confirm those 
found in \cite{machta97} and \cite{machta00}.  For both the 
Widom--Rowlinson model ($T=0$) and the Potts model (at least with this 
particular non-zero temperature) we found only a single phase transition  
point, and the onset of percolation is an order parameter for this transition.
As in these references, we also found that the transition is of second order  
for $q=2,3,4$, and of first order for $q\ge 5$. 
 
\subsection{Structure of pure phases} 
 
\begin{figure} 
  \begin{center} 
     { 
       \myfigure{height=13em}{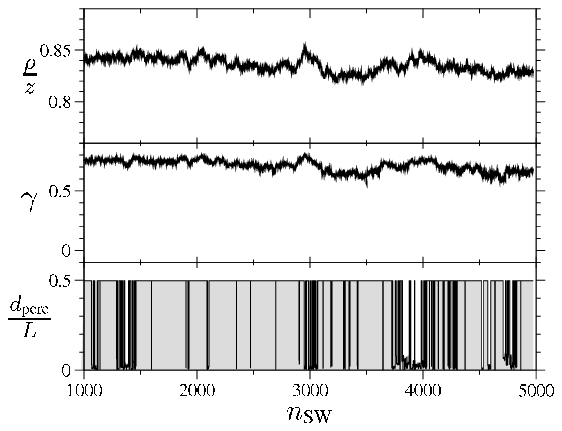} }\hfill 
     { 
      \myfigure{height=13em}{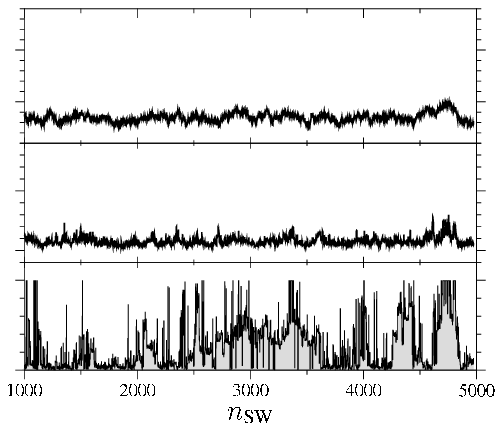} } 
     \caption{The evolution of the density, largest 
        cluster size, and percolation distance, 
       $\rho$, $\gamma$, and $\dperc $, respectively (see the text 
       for precise definitions).  Ordered initial conditions were used in 
       the left case, disordered in the right, while otherwise in both cases 
       $q=5$, $T=0$, $z=2.168\approx \zcrit$, and $L=512$.} 
    \label{fig:swq5} 
  \end{center} 
\end{figure}

\begin{figure} 
  \begin{center} 
     { 
      \myfigure{height=13em}{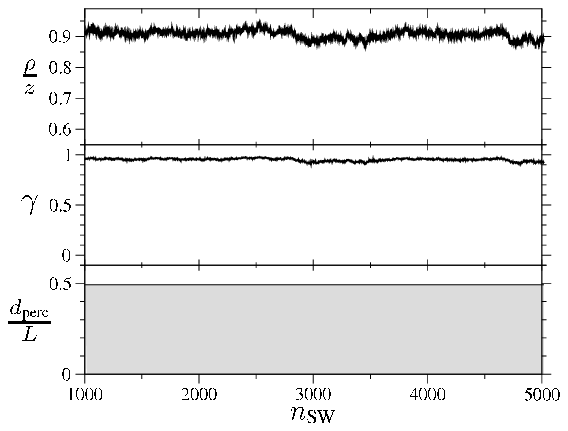} }\hfill 
     { 
        \myfigure{height=13em}{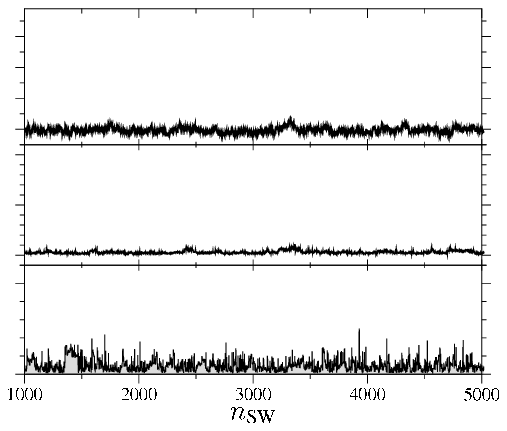} } 
     \caption{As in Figure \ref{fig:swq5}, but for the case 
        $q=10$, $T=0$, $z=2.56$, and $L=128$.} 
    \label{fig:swq10} 
  \end{center} 
\end{figure}

Apart from localizing the critical activity $z_c$ and clarifying the nature 
of the transition, our simulation measurements also provide some insight  
into the structure of the pure phases.  Let us start by considering Figure 
\ref{fig:swq5} which shows the evolution of our observables  $\rho$ 
(density), $\gamma$ (largest cluster size)  and $\dperc$ (percolation 
distance) for $q=5$ during the simulation steps for two  initial 
conditions: ordered (left-hand side) and disordered (right-hand side).  It 
is clear that $\rho $ and $\gamma$ display more or less stationary 
fluctuations  around some value that depends on the initial condition. This 
indicates the stability of  the ordered and disordered phases over the 
observation period and allows to infer that these phases coexist; recall 
the discussion in Section \ref{sec:absorption}. It also shows 
that the ordered phases have a higher particle density than the disordered 
phase, which means that the particle density in infinite volume should have 
a jump at $z_c$. Likewise, the proportion of particles in the largest 
cluster is nearly $1$ in the ordered case and nearly $0$ in the disordered 
case, from which we conclude that the typical configurations of the ordered 
phases contain a macroscopic cluster,  while those of the disordered phase 
do not. A glance at the evolution of $\dperc$ reveals that, in the ordered 
case, the macroscopic cluster typically hits the central $4\times 4$ square. 
On the other hand,  in the disordered case we see many ``spikes'', telling 
that it can quite well happen that one can walk a long distance from the  
central $4\times 4$ square along the random graph, but the corresponding 
clusters  are quite ``fragile'' and filamented, surviving only for a short 
time.  All these effects become more pronounced when $q$ gets larger, as 
can be seen from  Figure \ref{fig:swq10} for the case $q=10$. 
 
\begin{figure} 
  \begin{center} 
     { 
      \myfigure{height=13em}{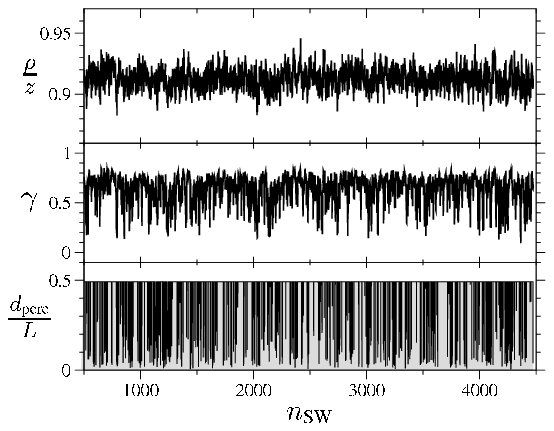} }\hfill 
     { 
        \myfigure{height=13em}{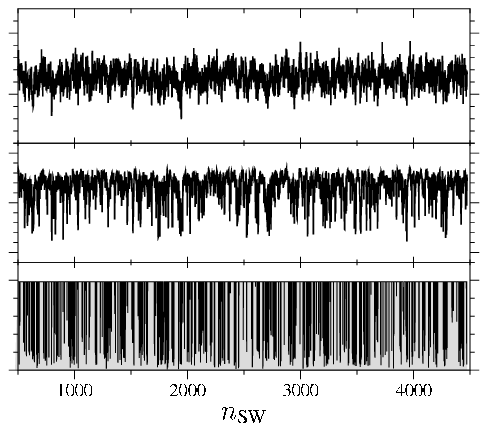} } 
     \caption{As in Figure \ref{fig:swq5}, but for the case 
        $q=2$, $T=0$, $z=1.72$, and $L=128$.} 
    \label{fig:swq2} 
  \end{center} 
\end{figure}

\begin{figure} 
  \begin{center} 
     { 
      \myfigure{height=13em}{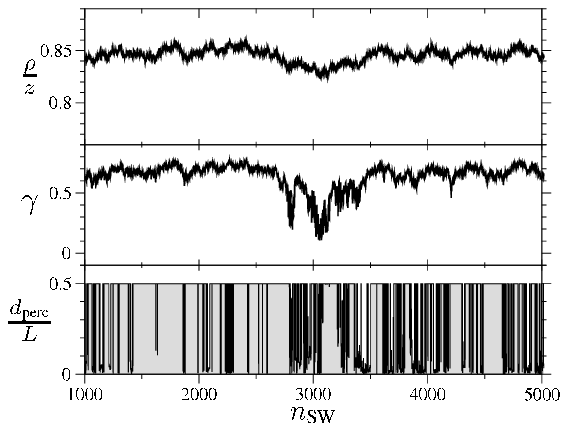} }\hfill 
     { 
        \myfigure{height=13em}{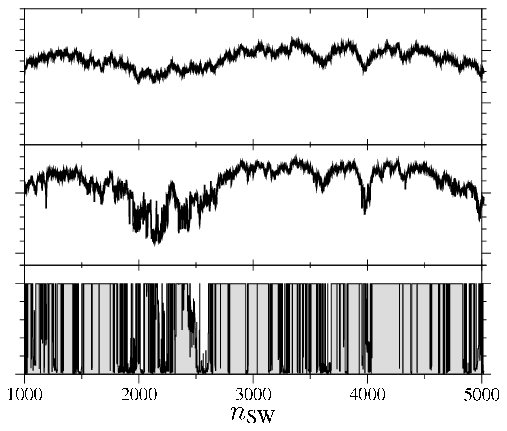} } 
     \caption{As in Figure \ref{fig:swq5}, but for the case 
        $q=4$, $T=0$, $z=2.05125$, and $L=512$.} 
    \label{fig:swq4} 
  \end{center} 
\end{figure}

By way of contrast, Figures \ref{fig:swq2} and \ref{fig:swq4} show the 
corresponding structure in cases $q=2$ and $q=4$. It is obvious that 
essentially no difference can be found between the ordered and disordered 
initial conditions, and that the criticality of $z_c$ manifests itself only 
by very large  fluctuations. We thus conclude that the phase transition is 
of second order.  In fact, it also becomes clear that $q=4$ is a boundary 
case: the portion of particles in the largest cluster is typically quite 
large, and so is the percolation  distance. This gives the hint that the  
onset of percolation at $z_c$ is quite rapid, and that the underlying 
value of $z$ in Figure \ref{fig:swq4} is actually slightly above $z_c$. 
 
\begin{figure} 
  \begin{center} 
     { 
       \myfigure{width=17em}{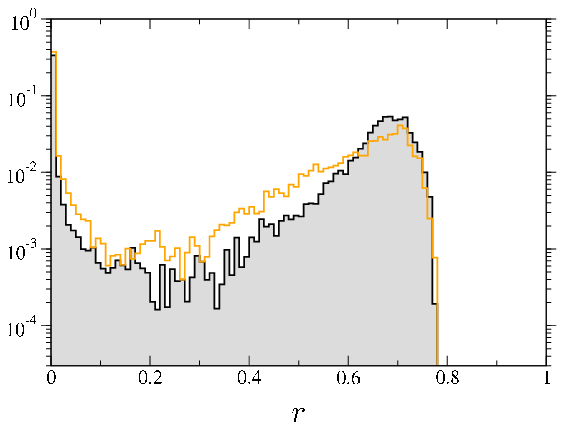} }\hfill 
     { 
       \myfigure{width=17em}{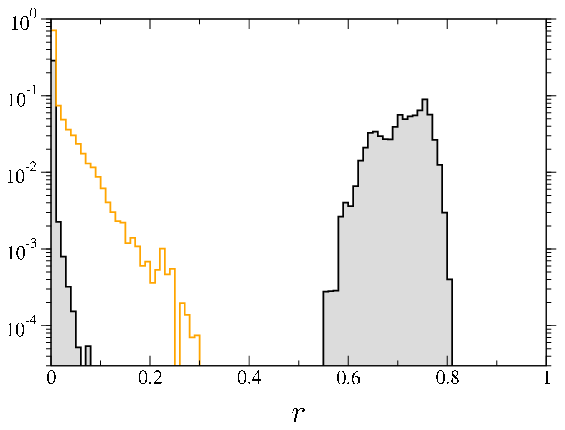} } 
     \caption{Histograms for the probabilities of finding a particle in 
       a cluster which contains a portion $r$ of the particles.  On the left, 
       the data were obtained from the runs with $q=4$ depicted in  
       Figure \ref{fig:swq4} with the shaded area giving the histograms for 
       the ordered initial condition and the gray line for the disordered.   
       The right figure shows the corresponding results for the  
       $q=5$ runs given in Figure \ref{fig:swq5}. } 
    \label{fig:histograms} 
  \end{center} 
\end{figure} 
 
\begin{figure} 
  \begin{center} 
    {  
      \myfigure{width=16em}{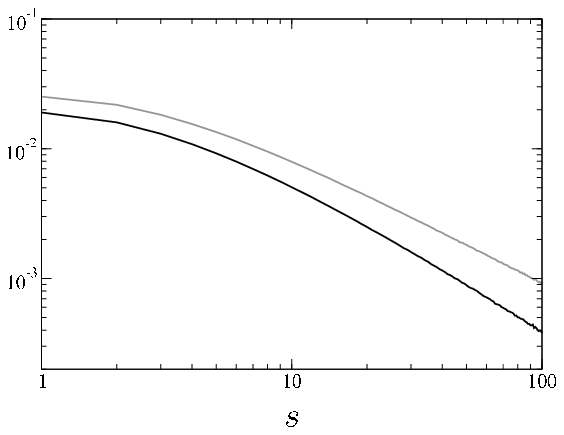} } 
    \caption{The probability of finding a particle in a cluster of size 
      $s$.  Measured from the same runs as the $q=5$ part of Figure 
      \ref{fig:histograms}, again black corresponding to the ordered initial 
      condition and grey to the disordered one.} 
    \label{fig:smallc_q5} 
  \end{center} 
\end{figure} 

Figure \ref{fig:histograms} presents the cluster-size distributions  
in the second-order case $q=4$ and the first-order case $q=5$, again for 
ordered and disordered initial conditions.  For $q=4$, there is again 
almost no influence of the initial conditions, while for $q=5$ there is a 
dramatic difference, and the different phases are separated by a range of 
values of cluster size which were not observed in either phase: those 
corresponding to the case when about half of the particles are in the 
maximal cluster.  Finally, Figure \ref{fig:smallc_q5} shows that in both 
phases the portion of particles in small clusters decays like a power-law 
of the cluster size (note the log-log scale in the figure).  However, in 
the ordered phase the decay is clearly faster.

\section*{Acknowledgments} 
 
We would like to thank J.~Machta for useful discussions and 
kindly providing us with details about their simulations. 
J.~Lukkarinen also wishes to acknowledge support from the Deutsche
Forschungsgemeinschaft (DFG) project SP~181/19-1.

\end{document}